\input{aipcheck}
 
\documentclass[
    ,final            
  ]
  {aipproc}

\usepackage{bm}
\usepackage{amsfonts,amssymb,amsmath}

\newcommand{\be}{\begin{equation}}
\newcommand{\bea}{\begin{eqnarray}}
\newcommand{\ee}{\end{equation}}
\newcommand{\eea}{\end{eqnarray}}
\newcommand{\bpi}{\begin{picture}}
\newcommand{\bce}{\begin{center}}
\newcommand{\epi}{\end{picture}}
\newcommand{\ece}{\end{center}}

\newcommand{\sla}{\slash \hspace{-0.22cm}}

\layoutstyle{8x11double}

\begin{document}

\title{Chiral symmetry breaking revisited: the gap equation with lattice ingredients}


\author{Arlene C. Aguilar}{
address={Federal University of ABC, CCNH, \\
Rua Santa Ad\'{e}lia 166, CEP 09210-170, Santo Andr\'{e}, Brazil.}
}

\begin{abstract}
 
We study chiral symmetry breaking in QCD, using 
as ingredients in the  quark gap equation  
recent lattice results for  the gluon  and ghost  propagators. 
The Ansatz employed for the quark-gluon vertex is purely non-Abelian, 
introducing a crucial dependence on the ghost dressing function and  
the quark-ghost scattering amplitude. The numerical impact 
of these quantities is considerable: the need to invoke confinement explicitly is avoided, and the  
dynamical quark masses generated are of the order of 300 MeV. 
In addition, the pion decay constant and the quark condensate are computed, 
and are found to be in good agreement with phenomenology. 

\end{abstract}

\maketitle


One of the major challenges of the strong interactions 
is to understand the underlying mechanism 
that generates masses for the  quarks and breaks the
chiral symmetry (CS). The CS breaking is an inherently nonperturbative
phenomena, whose study in the continuum
leads almost invariably to a treatment based
on the Schwinger-Dyson (SD) equation for the quark propagator
(gap equation).
As is well known to the SD experts, the existence or not
of  solutions for this equation depends crucially
on the details of its kernel, which is 
essentially composed by the fully dressed gluon propagator 
and the quark gluon  vertex~\cite{Aguilar:2010cn}. The latter quantity  controls  
the way that ghost  sector enters
into  the  gap  equation, and introduces  a
numerically crucial dependence on  the ghost dressing function~\cite{Fischer:2003rp} and 
the quark-ghost scattering amplitude.

In the present talk we report on a recent study of CS  breaking~\cite{Aguilar:2010cn} 
using the  SD equation for the quark propagator,
supplemented  with three nonpertubative ingredients: (i)  gluon propagator  and  (ii) ghost
dressing function obtained  from large-volume lattice simulations, and (iii)
the ``one-loop dressed''  approximate version of the scalar form factor of the 
quark-ghost scattering kernel.

The  starting
point is to express the fully dressed quark propagator
in the following general form~\cite{Roberts:1994dr}
\be
S^{-1}(p) = \sla{p} -m -\Sigma(p) = A(p^2)\,\sla{p} - B(p^2) \,,
\label{qpropAB}
\ee
where $m$  is the bare current quark mass, and  $\Sigma(p)$  the quark
self-energy. We consider the case without explicit CS breaking, i.e.,
bare mass m = 0. The dynamical quark mass function
can be defined  as being the ratio   
\mbox{${\mathcal{M}}(p^2)= B(p^2)/A(p^2)$}. Then
CS  breaking takes place when the scalar
component, $B(p^2)$  develops
a non-zero value.

\begin{figure}[!t]
\resizebox{0.9\columnwidth}{!}
{\includegraphics{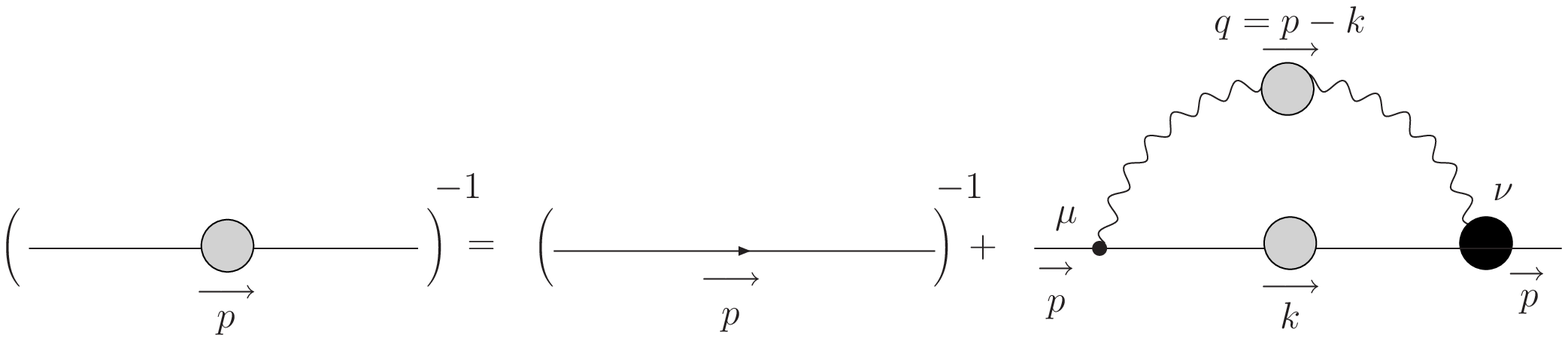}}
\caption{The quark SD equation (gap equation).}
\label{gap_eq}
\end{figure}

\begin{figure}[!b]
\resizebox{0.75\columnwidth}{!}
{\includegraphics{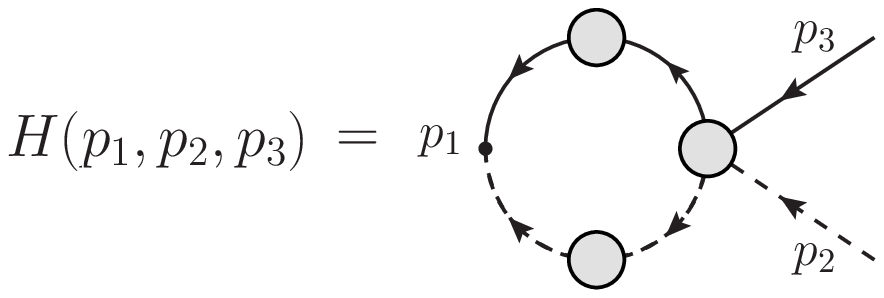}}
\caption{The quark-ghost scattering kernel.}
\label{figh}
\end{figure}

\begin{figure}[!h]
\begin{minipage}[b]{0.45\textwidth}
\centering
\includegraphics[scale=0.3]{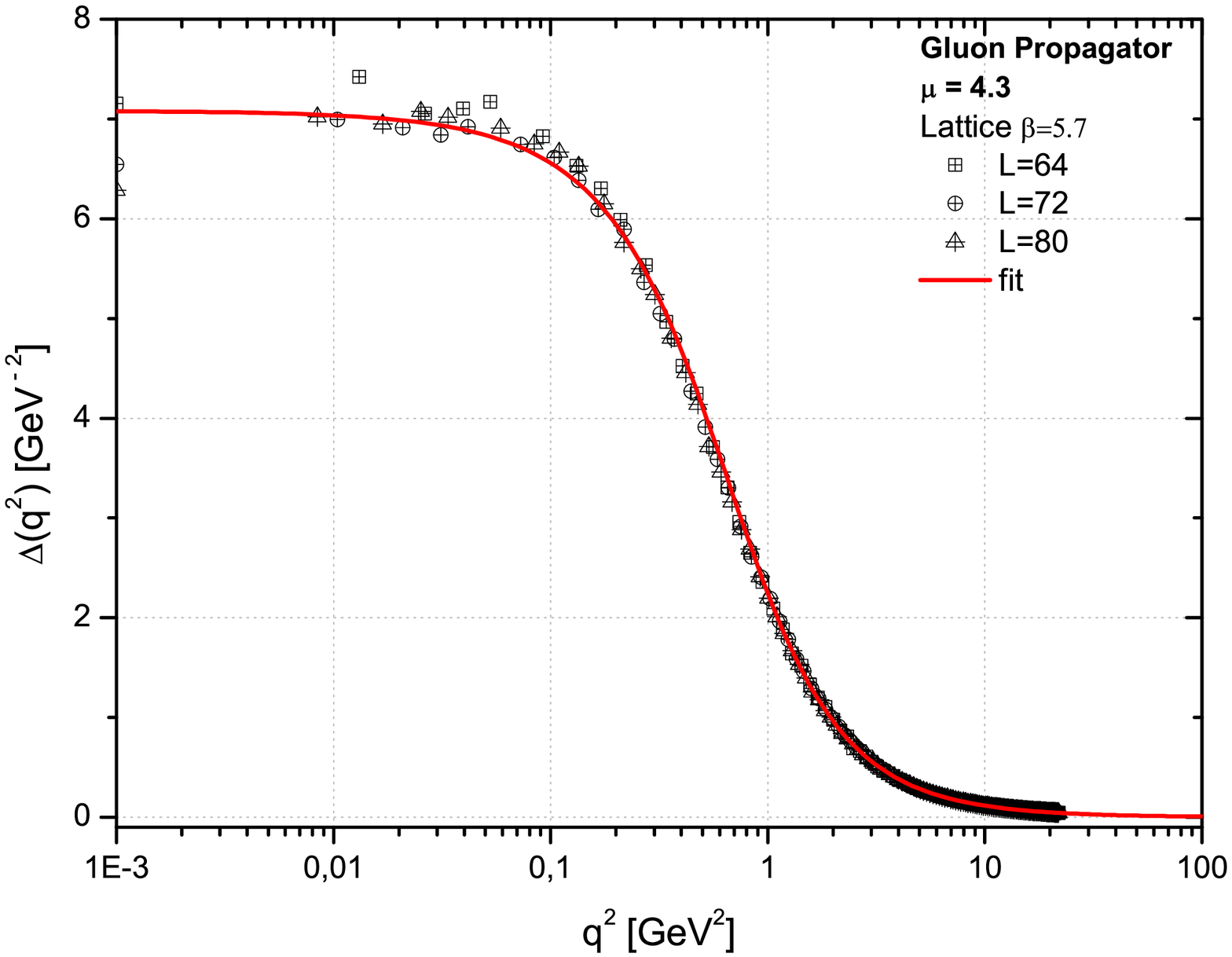}
\end{minipage}
\hspace{0.5cm}
\begin{minipage}[b]{0.50\textwidth}
\includegraphics[scale=0.3]{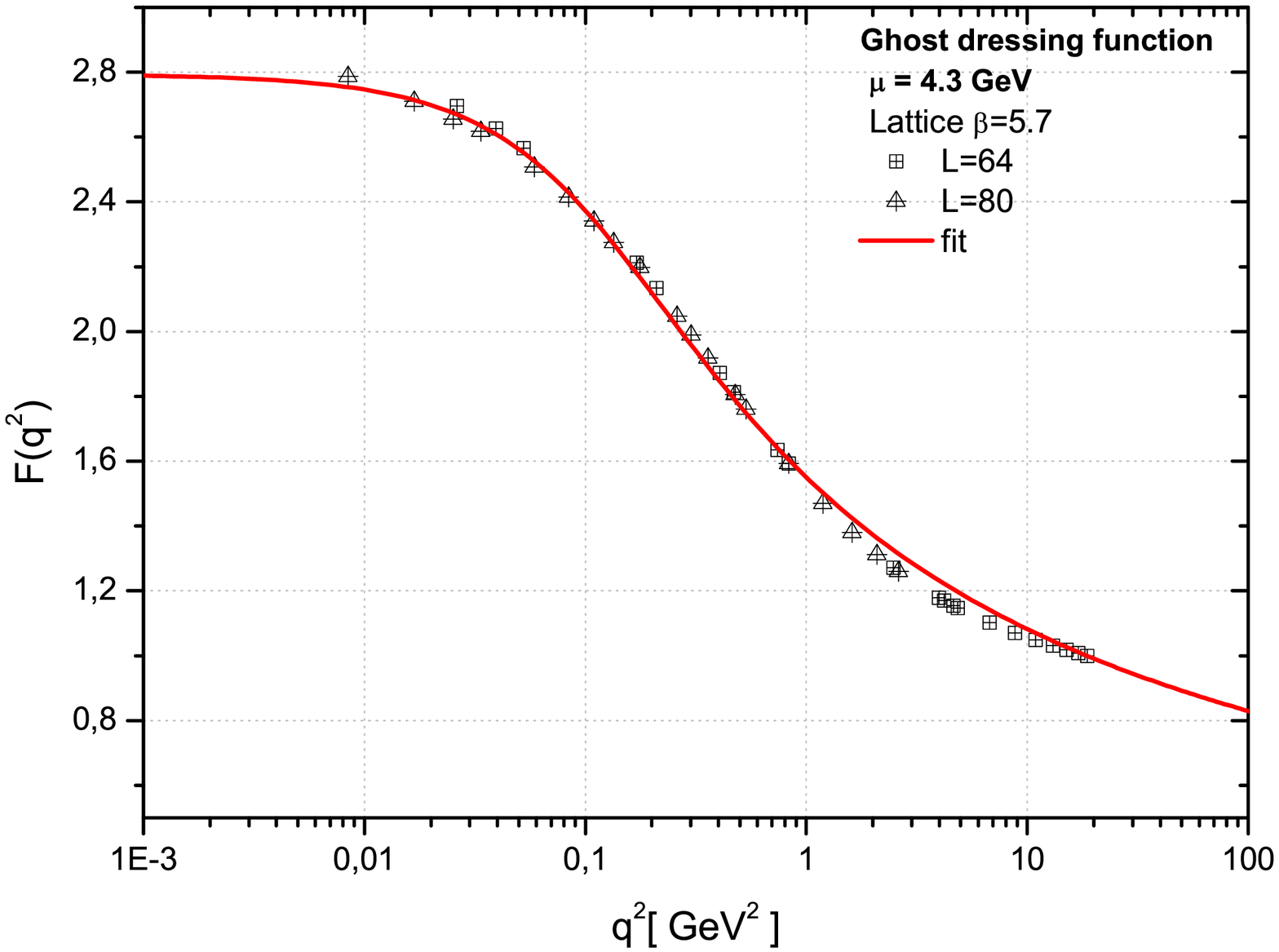}
\end{minipage}
\caption{Lattice results for the gluon propagator, $\Delta(q)$, and ghost dressing, $F(q)$ renormalized at $\mu=4.3$ GeV.}
\label{lattice}
\end{figure}
The SDE for the quark propagator, which is represented
schematically in Fig.~\ref{gap_eq}, can be written as
\begin{equation}
S^{-1}(p)= \sla{p} -C_{\rm F}g^2\int_k
\gamma_{\mu}S(k)\Gamma_{\nu}(-p,k,q)\Delta^{\mu\nu}(q) \,,
\label{senergy}
\end{equation}
where $q\equiv p-k$, \mbox{$\int_{k}\equiv\mu^{2\varepsilon}(2\pi)^{-d}\int\!d^d k$}, 
with \mbox{$d=4-\epsilon$} the dimension of space-time. $C_{\rm F}$  is the Casimir eigenvalue 
in the fundamental representation (for  $SU(3)$  \mbox{$C_{\rm F}=4/3$}). 
The full gluon propagator $\Delta_{\mu\nu}(q)$, in the Landau gauge,  has the form  
\be 
\Delta^{\mu\nu}(q)= -i\left[  g^{\mu\nu} - \frac{q^{\mu} q^{\nu}}{q^2}\right]\Delta(q^2) \,,
\label{fprop}
\ee
where the non-perturbative behavior of the scalar factor $\Delta(q^2)$ has been studied
in great detail in the continuum \cite{Aguilar:2006gr,Dudal:2008sp} and lattice simulations  
\cite{Bogolubsky:2007ud,Cucchieri:2007md}. The fully-dressed quark-gluon 
vertex $\Gamma_{\nu}(-p,k,q)$ also obeys its own  SD equation  which 
unfortunately, it is too complicated. Then, we have to resort to the so called gauge-technique,
where a nonperturbative Ansatz for the vertex $\Gamma_{\mu}(p_1,p_2,p_3)$  written in  
terms  of $S(p)$ is constructed, based on the requirement that it 
should satisfy the fundamental Slavnov-Taylor identity (STI)
\be
p_3^{\mu}\Gamma_{\mu} = 
F(p_3)[S^{-1}(-p_1) H - {\overline H} S^{-1}(p_2)]\,,
\label{STI}
\ee
where the ghost dressing function $F(p_3)$ is related to the full ghost propagator $D(p_3)$ 
by \mbox{$D(p_3)= iF(p_3)/p_3^2$}. The quark-ghost scattering kernel $H$, represented in 
Fig.~\ref{figh}, and its ``conjugated'' $\overline{H}$ are functions of the momenta $H=H(p_1,p_2,p_3)$, 
$\overline{H}=\overline{H}(p_2,p_1,p_3)$ respectively.

Both kernels $H$ and ${\overline H}$ have the 
following Lorentz decomposition~\cite{Davydychev:2000rt}
\bea
\hspace{-0.25cm}
H(p_1,p_2,p_3)\!\!\!\!\!\!\!\!\!\!&=&\!\!\!\!\!\!\!\!\!X_0 \mathbb{I}  
+X_1 \sla{p_1} +  
X_2  \sla{p_2} +
X_3 \tilde\sigma_{\mu\nu}p_1^{\mu} p_2^{\nu} \,,
\nonumber\\ 
\hspace{-0.25cm}
{\overline H}(p_2,p_1,p_3)\!\!\!\!\!\!\!\!\! &=&\!\!\!\!\!\!\!\!\! 
{\overline X}_0 \mathbb{I} 
-{\overline X}_2 \sla{p_1}  
-{\overline X}_1 \sla{p_2} 
+{\overline X}_3 \tilde\sigma_{\mu\nu}p_1^{\mu} p_2^{\nu},
\label{Xi}
\eea
where the form factors $X_i$ are functions of the momenta,
\mbox{$X_i=X_i(p_1,p_2,p_3)$},  
and we use the notation 
\mbox{${\overline X}_i (p_2,p_1,p_3) \equiv X_i (p_1,p_2,p_3)$}
and \mbox{$\tilde\sigma_{\mu\nu} \equiv \frac{1}{2}[\gamma_{\mu},\gamma_{\nu}]$}.

On the other hand, the most general Lorentz structure the longitudinal part of the vertex 
$\Gamma_{\mu}(p_1,p_2,p_3)$  appearing
in the lhs of Eq.~(\ref{STI}) is given by ~\cite{Davydychev:2000rt}
\bea
\Gamma_{\mu}(p_1,p_2,p_3) = 
  L_1 \gamma_{\mu}
+ L_2 (\sla{p_1} - \sla{p_2})(p_1-p_2)_{\mu} \nonumber\\ 
 \hspace{-0.35cm}+L_3 (p_1-p_2)_{\mu} 
+ L_4 \tilde\sigma_{\mu\nu}(p_1-p_2)^{\nu} \,,
\label{Li}
\eea
where once again we have suppressed the dependence on the momenta 
in the form factor $L_i$  [{\it i.e.} $L_i=L_i(p_1,p_2,p_3)$].Notice that, the 
tree level expression is 
recovered setting $L_1=1$ and $L_2=L_3=L_4=0$.

Due to the fact that the behavior of the vertex $\Gamma_{\mu}$ is constrained
by the STI of Eq.~(\ref{STI}), the form factors $L_i$'s appearing into
the Eq.~(\ref{Li}) will be given in terms of the form factors $X_i$'s of Eq.~(\ref{Xi}).

The full expressions for  $L_i$'s  in terms of the form factors $X_i$'s 
is given in \cite{Aguilar:2010cn}. For the sake of
simplicity, we will show here the case where only the scalar 
component of the quark-ghost scattering kernel is non-vanishing
{\it i.e.}  \mbox{$X_0 \neq 0$} while \mbox{$X_i = {\overline X}_i=0$}, 
for $i \leq 1$. In this limit, we obtain the following expressions 
\bea
L_1 &=& F(p_3) X_0(p_3)\left[\frac{A(p_1)+A(p_2)}{2}\right] \,,
\nonumber\\
L_2 &=& F(p_3) X_0(p_3)\left[\frac{A(p_1)- A(p_2)}{2(p_1^2 - p_2^2)}\right] \,,
\nonumber\\
L_3 &=& F(p_3) X_0(p_3)\left[\frac{B(p_1)- B(p_2)}{p_1^2 - p_2^2}\right] \,, 
\nonumber\\
L_4 &=& 0 \,.
\label{bci}
\eea

According the above expression the form factor $L_i$'s  displays an 
explicit dependence on the  product $F(p_3) X_0(p_3)$ which contains
information about the IR behavior of the ghost propagator. Therefore, 
the ghost sector couples to the  gap equation Eq.~(\ref{senergy})  through the 
quark-gluon vertex of  Eq.~(\ref{Li}). 
It is interesting to notice that in 
the limit of $F(p_3)=X_0(p_3)=1$ the form factors of Eq.~(\ref{bci}) reduces to the ones
used so-called Ball-Chiu (BC) vertex~\cite{Ball:1980ay}.

We will next insert into Eq.~(\ref{senergy})
the general quark-gluon vertex of Eq.~(\ref{Li}) with the 
expressions for the form factors $L_i$ given in Eq.~(\ref{bci}). Defining \mbox{$p_1=-p$}, \mbox{$p_2=k$}, and 
\mbox{$p_3=q$} and taking appropriate traces, we derive the expressions for the integral 
equations satisfied by $A(p^2)$ and $B(p^2)$ that schematically can be written as 
\bea
A(p^2)&=& 1 + C_{F}g^2  Z_c^{-1}\,\int_{k}\,
\frac{{\cal K}_0(p-k)}{A^2(k^2)k^2+B^2(k^2)}{\cal K}_A(k,p)\,, \nonumber\\ 
B(p^2) &=& C_{F}g^2  Z_c^{-1}\int_{k}\,\frac{{\cal K}_0(p-k)}{A^2(k^2)k^2+B^2(k^2)} {\cal K}_B(k,p)\,,
\label{scalar}
\eea
where the kernel ${\cal K}_0(q)$ corresponds to the part 
that is not altered by the tensorial structure of the quark-gluon vertex, namely
${\cal K}_0(q) =\Delta(q)F(q)X_0(q)$,
while the parts affected are ${\cal K}_A(k,p)$ and ${\cal K}_B(k,p)$. 

The gap equation depends  on the nonperturbative form 
of the three basic Green's functions, namely $\Delta(q)$, $F(q)$, and $X_0(q)$.
For $\Delta(q)$ and $F(q)$ we use the recent lattice data obtained by ~\cite{Bogolubsky:2007ud}, and
shown in Fig.~\ref{lattice}.

\begin{figure}[!t]
\resizebox{0.8\columnwidth}{!}
{\includegraphics{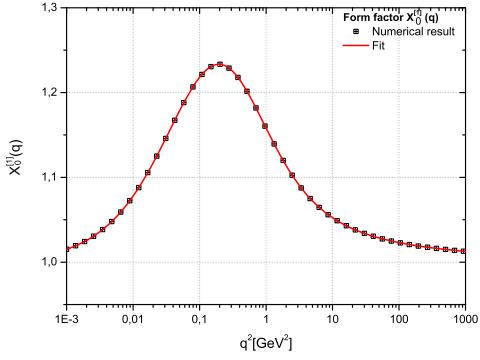}}
\caption{The form factor $X_0^{[1]}(q)$ given by Eq.~(\ref{sk2}.}
\label{sk}
\vspace{-0.75cm}
\end{figure}

We clearly see that both lattice results for $\Delta(q)$ and $F(q)$
are infrared finite. Such a
feature can be associated to a purely non-perturbative effect
that gives rise to a dynamical gluon mass \cite{Cornwall:1982zr}, which saturates 
the gluon propagator in the IR. The appearance of 
the gluon mass is also responsible for the infrared finiteness of the  
ghost dressing function, $F(q^2)$ \cite{Aguilar:2006gr,Boucaud:2008ji}, which is shown on the right panel
of Fig.~\ref{lattice}, 

Unfortunately for $X_0(q)$ there is no lattice data available, and  in order to 
obtain a non-perturbative estimate for $X_0$, we will study 
``one-loop dressed'' scalar contribution of the diagram of Fig.~\ref{figh}
 in an approximate kinematic configuration, which simplifies the 
resulting structures considerably. Specifically, we will assume that 
$p_1 = p_2 \equiv p$, and  $p= - q/2$. In doing so, we arrive at
(see details in \cite{Aguilar:2010cn})
\be
X_0^{[1]}(q) = 1 + \frac{1}{4} C_A g^2q^2 \int_k 
\left[1-\frac{(k \cdot q)^2}{k^2 q^2}\right]\Delta (k) F(k) \frac{F(k+q)}{(k+q)^4} \,,
\label{sk2}
\ee
We proceed substituting the fit for  the lattice data for $\Delta(q)$ and $F(q)$ presented in
Fig.~\ref{lattice} into Eq.~(\ref{sk2}). The  numerical result for $X_0^{[1]}(q)$ is shown in the  
Fig.~\ref{sk}.
%
\begin{figure}[!h]
\resizebox{0.8\columnwidth}{!}
{\includegraphics{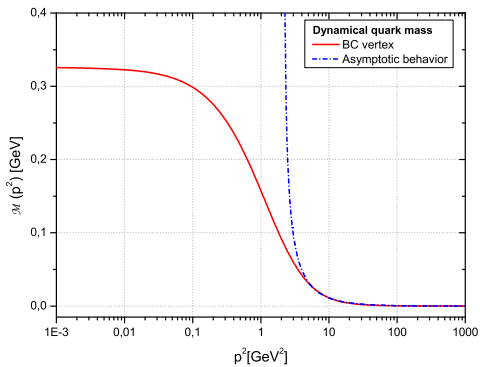}}
\caption{The quark mass ${\mathcal M}(p^2)$ for the BC vertex.}
\label{mass}
\vspace{-0.75cm}
\end{figure}

$X_0^{[1]}(q)$ shows a maximum  
located in the intermediate momentum region (around \mbox{$450$ MeV}),   
while in the UV and IR regions $X_0^{[1]}(q) \to 1$.
Although this peak is not very pronounced,  
it is  essential for providing to the kernel of the gap 
equation the enhancement required for the generation of 
phenomenologically compatible constituent quark masses. 

Now we are in position to solve the system formed by Eq.(\ref{scalar}) 
Substituting  $\Delta(q^2)$, $F(q^2)$, and $X_0^{[1]}(q)$ to  Eq.(\ref{scalar}), 
with the modification \mbox{$Z_c^{-1} {\cal K}_{A,B} (k,p) \to  {\cal K}_{A,B} (k,p) F(p^2)$},
to enforce the correct renormalization group behavior of the dynamical mass (see discussion in \cite{Aguilar:2010cn}), 
we determine numerically the unknown functions $A(p^2)$ and $B(p^2)$. The result
for the dynamical quark mass ${\mathcal M}(p^2)$ is shown in Fig.~\ref{mass}. 

One clearly sees that ${\mathcal M}(p^2)$ 
freezes out and acquires a finite value in the IR, \mbox{${\mathcal M}(0) = 294 $ MeV}. In the UV it shows the expected 
perturbative behavior represented
by the blue dashed curve.

With  the behavior of the dynamical quark mass at hand, we have computed 
pion decay constant and the quark condensate and we obtained 
\mbox{$f_{\pi}= 80.6$ MeV} and \mbox{$\left\langle \bar{q}q\right\rangle (1\,\mbox{GeV}^2)=\,(217 \mbox{MeV})^3$}
respectively, which are in good agreement with phenomenological results.

\vspace{-0.25cm}

\paragraph{Acknowledgments:
The author thanks the organizers QCHS-IX for the pleasant conference. 
This research is supported by the Brazilian Funding Agency CNPq under the grant 305850/2009-1
and 453118/2010-0.
\vspace{-0.25cm}}

\end{document}